\documentclass{cupconf}
\usepackage{natbib,graphicx}

\def\lesssim{\mathrel{\hbox{\rlap{\hbox{\lower4pt\hbox{$\sim$}}}\hbox{$<$}}}}
\def\gtrsim{\mathrel{\hbox{\rlap{\hbox{\lower4pt\hbox{$\sim$}}}\hbox{$>$}}}}

\def\macc {\dot M_*}

\def\ms {M_*}
\def\solmas{{M$_\odot$}}

\let\la=\lesssim
\let\ga=\gtrsim

\newcommand{\msun}{M_{\odot}}
\newcommand{\lsun}{L_{\odot}}

\newcommand{\sfrff}{\mbox{SFR}_{\rm ff}}

\newcommand{\apj}{ApJ}
\newcommand{\apjs}{ApJS}
\newcommand{\apjl}{ApJ}
\newcommand{\aap}{A\&A}
\newcommand{\mnras}{MNRAS}

\newcommand{\nat}{Nature}
\newcommand{\araa}{ARA\&A}
\newcommand{\aj}{AJ}
\newcommand{\pasj}{PASJ}

\title[Formation of Massive Stars]{Models for the Formation of Massive Stars}

\author[M.~R. Krumholz \& I.~A. Bonnell]{
M\ls A\ls R\ls K\ns R.\ns K\ls R\ls U\ls M\ls H\ls O\ls L\ls Z\ls$^1$\footnote{Hubble Fellow} and
I\ls A\ls N\ns A.\ns B\ls O\ls N\ls N\ls E\ls L\ls L\ls$^2$}

\affiliation{$^1$ Department of Astrophysical Sciences, Princeton University, Princeton, NJ 08544-1001\\
$2$ SUPA, School of Physics and Astronomy, University of St Andrews, KY16 9SS, UK}

\begin{document}

\maketitle

\begin{abstract}
The formation of massive stars is currently an unsolved problems in astrophysics. Understanding the formation of massive stars is essential because they dominate the luminous, kinematic, and chemical output of stars. Furthermore, their feedback is likely to play a dominant role in the evolution of molecular clouds and any subsequent star formation therein. Although
significant progress has been made observationally and theoretically, we still do not
have a consensus as to how massive stars form. There are two contending models to
explain the formation of massive stars, Core Accretion and Competitive Accretion. They differ
primarily in how and when the mass that ultimately makes up the massive star is gathered. In the
core accretion model, the mass is gathered in a prestellar stage due to the overlying pressure of a 
stellar cluster or a massive pre-cluster cloud clump. In contrast, competitive accretion envisions
that the mass is gathered during the star formation process itself, being funneled to the centre
of a stellar cluster by the gravitational potential of the stellar cluster. Although these differences
may not appear overly significant, they involve significant differences in terms of the
physical processes involved. Furthermore, the differences also have important implications in terms
of the  evolutionary
phases of massive star formation, and ultimately that of  stellar clusters and star formation on larger scales. Here we review the dominant models, and discuss prospects for developing a better understanding of massive star formation in the future.
\end{abstract}

\firstsection
\section{Introduction}

The difficulties in understanding massive star formation lie  in that we are not able to fully ascertain the 
properties of a cloud in which massive stars form or to determine whether the properties we do observe
are that of the pristine initial conditions or those due to subsequent evolution. For example, the 
central condensation often observed in massive cores could be an indication of the central
condensation required in the core accretion model, or equally represent the evolved state
from a dynamical collapsing model. It is also equally difficult to determine the census of young stars in these cores and to what degree their dynamics may be thus affected. We are therefore left
in a state of some ambiguity in terms of constraining  the initial conditions for the models.

There is also considerable ambiguity regarding the end state of the massive star formation process, particularly as regards the multiplicity and clustering of massive stars. Observations unambiguously show that massive stars have a much higher companion fraction that low mass stars \citep{duchene01, preibisch01, shatsky02, lada06}, even at early, still-embedded stages of star formation \citep{apai07}. A significant fraction of massive stars appear to be twins, systems with mass ratios near unity \citep{pinsonneault06}. However, the full mass and period distribution of massive stars is much less well-determined than it is for low mass stars, and the statistical significance of the \citeauthor{pinsonneault06} results has been disputed due to the small sample size and the potential for selection bias produced by using a sample of eclipsing binaries \citep{lucy06}.

Similarly, the clustering properties of massive stars are debated. While the great majority of massive stars are either part of star clusters or are runaways ejected from clusters, since on statistical grounds on would expect most massive stars to be in clusters it is unclear if this observation implies that O stars form only, or preferentially, in clusters. Statistical analyses of cluster data reach contradictory conclusions, with some finding that the stellar IMF is truncated at a value that is an increasing function of the mass of the cluster in which that star is found \citep{weidner04, weidner06}, and others concluding that the data are consistent with random and independent sampling from stellar and cluster mass functions, with a cluster mass function that is continuous down to a single star \citep{oey04, elmegreen06, parker07}. Observationally, $4\pm 2\%$ of O stars do not appear to be surrounded by clusters and are not runaways \citep{dewit04, dewit05}, which would seem to imply that O stars can form in isolation. However, it remains possible that these O stars formed via a different mechanism than most massive stars.


On the theoretical side, a traditional problem in massive star formation is how the mass is actually accreted by the proto-massive star and how this accretion circumvents the radiation pressure produced by the high stellar luminosity and the opacity of dust grains in the accretion flow. A number of potential solutions have recently been advanced based on the physics of disk accretion and ways of circumventing the effects of the photon pressure such that we can conclude massive star formation is unlikely to be halted solely due to radiation pressure. However, these conclusions are still preliminary, since to date no simulation of massive star formation including radiative effects has successfully demonstrated the formation of stars up to $\sim 80$ $\msun$, the mass of the largest star whose mass is securely measured \citep[the eclipsing spectroscopic binary WR 20a;][]{bonanos04, rauw05}.

In this review we begin by describing the two primary models of massive star formation in \S\S~\ref{coreaccretion}-\ref{compaccretion}, and we then focus on the challenges that both models face in \S~\ref{challenges}. We then suggest directions in which we can make progress on those challenges, and in distinguishing between the two models, in \S~\ref{future}.

\section{The Core Accretion Model}
\label{coreaccretion}

\subsection{The Model}

The basic premise of the core accretion model is that all stars, low and high mass, form by a top-down fragmentation process in which a molecular cloud breaks up into smaller and smaller pieces under the combined influence of turbulence, magnetic fields, and self-gravity. This process continues down to some smallest scale, called a core, which does not undergo significant internal sub-fragmentation before collapse to stellar densities. Thus a core represents the object out of which an individual star, or a bound stellar system (for example a binary) forms, and the mass of the core determines the mass reservoir available to form the star. A massive star must therefore form from a massive core \citep{mckee02, mckee03}. Cores also represent the scale in the star formation process at which gas becomes unstable to a global gravitational collapse. On all larger scales objects have gravitationally unstable parts within them, but they are not in a state of overall collapse, and they turn a relatively small fraction of their mass into stars per dynamical time. 

This model leads immediately to several predictions that may be tested against observations. First, in the core accretion scenario a core is the mass reservoir available for accretion onto the star that forms within it, so there must be a direct relationship between the core and stellar mass functions. Indeed, the core mass distribution will effectively set the stellar mass distribution. Second, for systems too young for the stars to have moved significantly from their birth sites, the spatial and velocity distributions of cores and stars should be similar -- properties such as clustering strength, degree of mass segregation, and kinematics should be similar for cores and for the youngest stars. Third, since the individual cores are gravitationally collapsing but the bulk of the gas is not, the overall rate of star formation in clouds much larger than an individual core should be small, in the sense that $\la 10\%$ of the mass should be   converted into stars per cloud dynamical time.

Finally, it is worth noting that the origin of the core mass function and core kinematics need not be specified as part of the core accretion model. Although a number of theories have been advanced to explain these quantities \citep[e.g.][]{padoan02}, these models and the core accretion model are independent of one another, and must be tested separately.

\subsection{Observational Tests}

\subsubsection{Core and Stellar Mass Functions}

Thus far the predictions of the core accretion model appear to hold up well against observations, although there are several significant caveats and areas of uncertainty. Observations of many different star-forming regions in dust continuum emission  \citep{motte98, testi98, johnstone01, onishi02, reid05, reid06a} and near infrared extinction mapping \citep{alves07} essentially all find that the core mass distribution has a functional form strikingly similar to the stellar IMF. The core mass function has a powerlaw tail of Salpeter slope at high masses, and a flattening at low masses. The most recent and sensitive study, that of \citet{alves07}, even hints at a turn-down in the mass function below the peak, exactly as is seen in the stellar IMF. The sole significant difference between the core and star mass functions appears to be that the core mass function is shifted to higher masses by a factor of $\sim 3$, exactly what one would expect if a significant fraction of pre-stellar cores were ejected by outflows rather than accreting onto a star \citep{matzner00}.

There is a significant caveat that the core mass function has only been reliably measured up to masses $\la 20$ $\msun$ on the Salpeter side, and down to $\ga 0.5$ $\msun$ on the brown dwarf side. The latter limitation comes from the difficulty of detecting such low mass objects, and the former comes from the difficulty of resolving objects at the large distances to which observations must reach to find such massive cores. While there are observations of cores up to a few hundred $\msun$, massive enough to form the largest stars \citep{beuther04b,sridharan05,beuther05b,beuther06a, reid06a}, these objects are both expected and observed to have internal structure, and so one might worry that they are simply blended collections of low mass cores. High resolution interferometric observations show that at least some of these massive cores are very strongly centrally concentrated, however, which argues against this possibility \citep{beuther07b}. That said, it cannot be ruled out definitively without higher resolution observations or a larger sample of massive cores.

\subsubsection{Core Spatial and Kinematic Distributions}

The prediction of the core accretion model that young stars and star-forming cores should have similar spatial distributions also appears to be consistent with observations. In young clusters, stars with masses $\ga 5$ $\msun$ are preferentially located in cluster centers, but at lower stellar  masses the mass function appears to be independent of position within a cluster \citep[e.g.][]{hillenbrand98}. There is some dispute about whether this segregation is a function of birthplace \citep{bonnell98b, huff06} or is produced dynamically in young clusters \citep{tan06a, mcmillan07}, but to the extent that it is primordial it should be reflected in the spatial distribution of cores. In at least one star-forming cloud, $\rho$ Ophiuchus, exactly such a pattern of core segregation is observed: massive cores are found only in the center, but cores $\la$ 5 $\msun$ show a position-independent mass function \citep{elmegreen01, stanke06}. The major caveat regarding this observation is that it is not been replicated yet in more distant and more massive star-forming clouds, so it is unclear to what extent it is a generic feature of massive star-forming regions.

On the question of kinematics, observations also appear consistent with the core accretion model. Observations of populations of pre-stellar cores and cores around class 0 protostars \citep[e.g.][]{goodman98, walsh04, andre07, kirk07, rathborne07} consistently find four consistent kinematic signatures for low mass cores across a wide range of cluster-forming clouds: first, the linewidths along sightlines that pass through core centers are essentially always sonic, with no significant turbulent component. Second, the linewidth increases only slowly as one examines sightlines progressively further from core centers, becoming only transonic (i.e.\ Mach numbers $\sim 2$) at projected distances of $\sim 0.1$ pc from core centers. Third, the mean velocity difference between cores and their envelopes are also small. Fourth, if one computes a centroid velocity for each core, the dispersion of core centroid velocities is generally smaller than the virial velocity in the observed region. 

Simulations of cores forming in the context of driven, virialized turbulent flows in which the large-scale region is not globally collapsing are able to reproduce most but not all of these observations, while simulations in which the region is in global collapse have much greater difficulty \citep{offner07a}. In non-collapsing regions cores have subsonic or transonic velocity dispersions along the line of sight because the collapse is limited and localized, although the linewidths are still somewhat larger than observed ones, probably due to the absence of magnetic fields in the simulations. Because collapse is localized, cores are not bounded by strong infall shocks and the velocity dispersion at their edges is fairly small. Cores move sub-virially with respect to one another because they are born from the densest, shocked gas that is preferentially at low velocity dispersion \citep{padoan01, elmegreen07}.

In contrast, simulations in which the turbulence is not driven and the entire region enters a process of global collapse have much greater difficulty \citep[e.g.][]{klessen05,ayliffe07}. These simulations also find relatively low differences in mean velocity between cores and envelopes, but, due to the large-scale infall that occurs in them, they generally also find supersonic velocity dispersions, either at core centers or in their outskirts. They also produce core-to-core velocity dispersions that are about the same size as those seen in the gas \citep{offner07a}, rather than smaller as the observations demand. This feature occurs because the cores are effectively dissipationless, and retain the velocity dispersions they have at birth. The gas, however, continually loses energy through shocks, and thus after a while its velocity dispersion decreases to the point where it matches that of the cores.

\subsubsection{The Star Formation Rate}

The final test of the core accretion model is the overall rate of star formation. Core accretion requires that, at any instant, collapse in cloud be confined to discrete regions containing a small fraction of the mass. Thus, the star formation rate per cloud dynamical time must be small, $\la 10\%$. In contrast, competitive accretion models require that cloud convert their mass to stars quickly, with a star formation rate per dynamical time $\ga 10\%$, at least by the end of the star formation process or when averaged over many clouds.

This prediction can be tested in two ways, one statistical and one object-by-object. Statistically, one can estimate star formation rates in a large sample of clouds by observing a galaxy in a tracer of star formation such as infrared emission, and then compare it to the total mass of dense, cluster-forming gas by observing the same galaxy in a dense gas tracer such as the HCN($1\rightarrow 0$) line, which is emitted by gas at densities $\sim 10^5$ cm$^{-3}$ \citep{gao04b, gao04a, wu05}. If the dense gas mass divided by the free-fall time at the mean density of that gas is much larger than the measured star formation rate in the galaxy, then the clouds must, on average, convert their mass into gas relatively slowly. In contrast, if the cloud mass divided by the free-fall time is comparable to the galactic star formation rate, then the clouds must convert into stars rapidly.

\citet{krumholz07e} perform this exercise using a variety of tracers of dense gas, and find that, with the exception of one tracer that only yields an upper limit, all the data require that clouds convert to stars at a rate well under $10\%$ of the mass per dynamical time. This appears to be consistent with the requirements of core accretion. However, it is worth mentioning the caveat that this technique averages over a large number of star-forming clouds and assumes that all of them are active star-formers. If there are a significant number of clouds that do not form stars, then the star formation rate in the active star-formers could be significantly higher \citep{elmegreen07}. This picture would be inconsistent with core accretion, although it would also demand an explanation for what mechanism prevents gas at densities of $10^5$ cm$^{-3}$ from forming stars.

For individual objects, one can estimate the star formation rate by computing the age spread within a star cluster, and then estimating the efficiency of conversion of gas into stars based on cluster dynamics. The age divided by the dynamical time of the parent cloud multiplied by the efficiency yields the fraction of the mass that must have been converted into stars per dynamical time. This method is considerably less certain, since for clusters where it is possible to estimate an age spread using pre-main sequence tracks the parent cloud has generally already been dispersed. One must therefore estimate the mass of that cloud (in order to determine the overall efficiency) and its dynamical time without observing it directly. Moreover, even the age spreads themselves are uncertain, since the pre-main sequence tracks used to age-date stars are of limited accuracy.

Given these uncertainties, it is not surprising that interpretations of the data vary widely. \citet{elmegreen00, elmegreen07} argues that star formation in a cluster generally ends in only $1-2$ crossing times, which, together with a typical cluster formation efficiency of $20-50\%$ \citep{lada03}, implies rapid star formation. \citet{tan06a} and \citet{huff06}, analyzing much the same data set, conclude instead that the typical duration of star formation is $4-5$ crossing times, implying a much lower rate of star formation. Resolving this controversy will require both better observations of still partially embedded young clusters and improvements in the pre-main sequence models.

It is worth noting, however, that several groups recently obtained parallax measurements of the distance Orion Nebula Cluster, the object about which much of the debate regarding the duration of cluster formation has taken place. These observations correct the distance from roughly $500$ pc to about $400$ pc \citep{sandstrom07, caballero07, hirota07, menten07}, which will reduce the inferred luminosity for all the Orion stars by $\sim 30\%$. This will in turn increase the age spread inferred from the pre-main sequence tracks. This revision obviously favors the extended formation hypothesis, although determining by exactly how much will have to await a full re-analysis of the locations of ONC stars on pre-main sequence tracks using the revised luminosities. 

\subsection{Massive Stars in the Core Accretion Model}

\subsubsection{Initial Fragmentation}

In the context of the core accretion model, a massive star forms from a core that is similar to those that form low mass stars. The one challenge that is particular to the core accretion model is how a massive star, an object that is potentially hundreds of Jeans masses in size, can form by direct collapse. Why does the fragmentation process not always continue down to objects that are $\sim 1$ Jeans mass in size? Indeed, purely hydrodynamic simulations of core collapse and fragmentation find exactly this behavior \citep{dobbs05}.

\begin{figure}
\centerline{\includegraphics{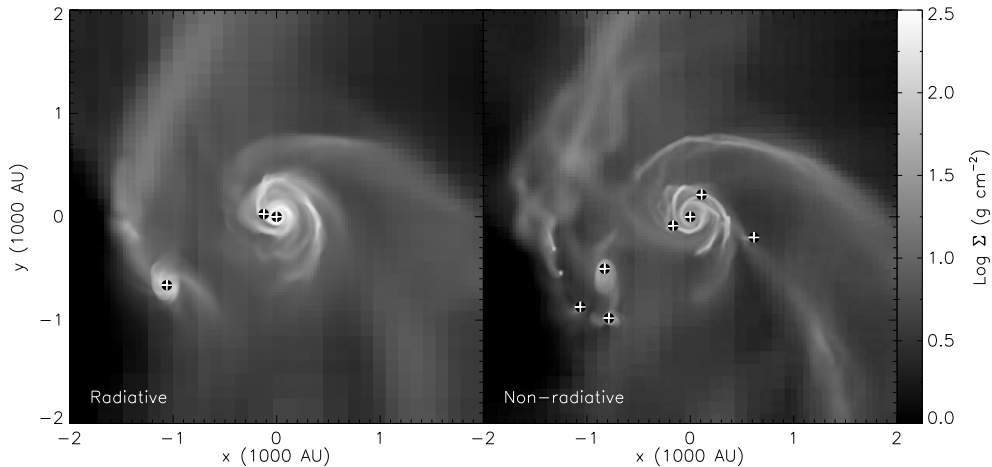}}
\caption{
\label{radfragment}
The plots show the column density through a $4000$ AU around the most massive protostars in two simulations of the collapse and fragmentation of a massive core by \citet{krumholz07a}. The simulation shown on the left includes radiative transfer and feedback, while the one on the right omits these effects, but they are otherwise identical in initial conditions, evolutionary time, and resolution. The plus signs indicate the locations of protostars.
}
\end{figure}

The simulations and analytic work of \citet{krumholz06b} and \citet{krumholz07a, krumholz07b} provide a likely answer: at very early times in the star formation process, radiation feedback powered by the gravitational potential energy of collapsing gas will modify the effective equation of state of the gas on scales of $\sim 0.01-0.1$ pc, roughly the size of observed pre-stellar cores. The effective equation of state is critical to determining whether or not gas fragments \citep{larson05, jappsen05}: equations of state that are isothermal or softer favor fragmenation, while stiffer equations of state prevent it. In the simulations of \citeauthor{krumholz07a}, the accretion luminosity coming from the first $\sim 0.1-1$ $\msun$ star in the simulation gives the gas an equation of state with an effective ratio of specific heats $\gamma=1.1-1.2$, stiffer than isothermal and thus unfavorable to fragmentation. This not truly an equation of state, since the gas temperature depends on position relative to the illuminating source as well as density, but it appears to have the same effect on fragmentation as a true equation of state. In simulations with radiation, \citeauthor{krumholz07a} find that a typical massive core forms only a handful of fragments, and most of these form out of gravitationally unstable protostellar disks that are shielded from radiation by their high column densities. Figure \ref{radfragment} illustrates this result, showing a comparison between a non-radiative and a radiative simulation starting from identical initial conditions. The strong suppression of fragmentation found in these simulations suggests that in the dense regions where massive stars form gas {\it cannot} fragment on size scales below a few hundreths of a parsec, so monolithic collapse is the only possibility.

The major remaining caveat on this work is that the simulations begin from somewhat unrealistic initial conditions in which cores possess turbulent velocity structures but not turbulent density structures. In reality, even at the time when a first protostar forms there should be density structure present in the gas, and some overdense regions may be sufficiently shielded against radiation to collapse despite significant external illumination. Future simulations using radiative transfer on larger scales will have to examine this effect.

\subsubsection{Massive Protostellar Disks and Companion Formation}

Since the collapsing core has non-zero angular momentum, a protostellar disk naturally results. The sizes of these disks are typically $\sim 1000$ AU, a value imposed by the core size and the typical angular momentum carried by cores' turbulent velocity fields. These disks are quite different from those around lower mass protostars in low density regions. Due to the high density in massive cores, the accretion rate onto these disks is extremely high. Early in their lifetimes rapid accretion drives massive protostellar disks to a regime of gravitational instability characterized by Toomre $Q$ values near unity and disk to star mass ratios $\sim 0.5$ \citep{kratter06, krumholz07a, kratter07}. While in this stage the disks have strong spiral arms and are subject to gravitational fragmentation. This condition persists throughout the majority of the time during which the massive protostar is accreting.

Disks in this evolutionary phase are also subject to fragmentation. Fragments that form in the disk in tend to migrate inward as the disk accretes. The fate of these fragments is unknown, and most of them likely merge with the central protostar. However, it seems likely that at least some of them survive to become close companions, explaining the high companion fraction of massive stars. Those which are come closer than $\sim 1$ AU are likely to gain considerable mass via mass transfer from the primary, which passes through a phase of rapid expansion to large radii on its way to the main sequence. Such systems are likely to become massive twins \citep{krumholz07c}.

\begin{figure}
\begin{center}
\includegraphics[scale=0.85]{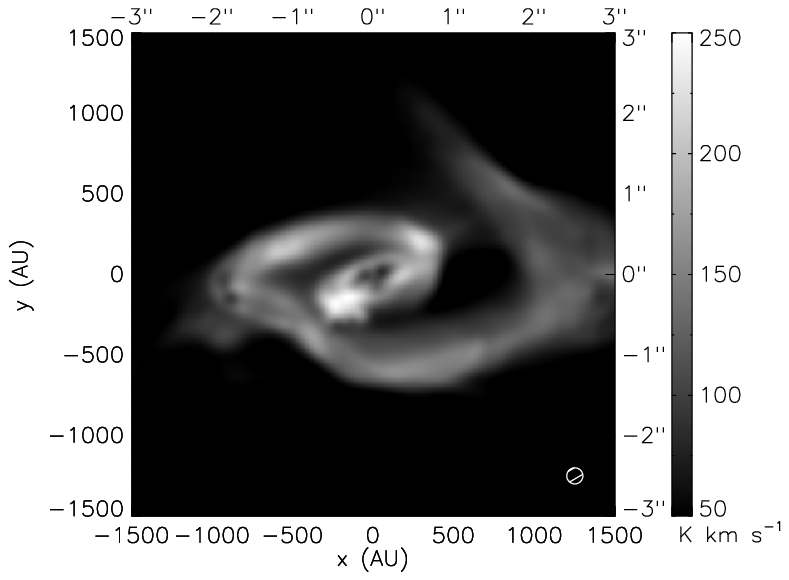}\includegraphics[scale=0.85]{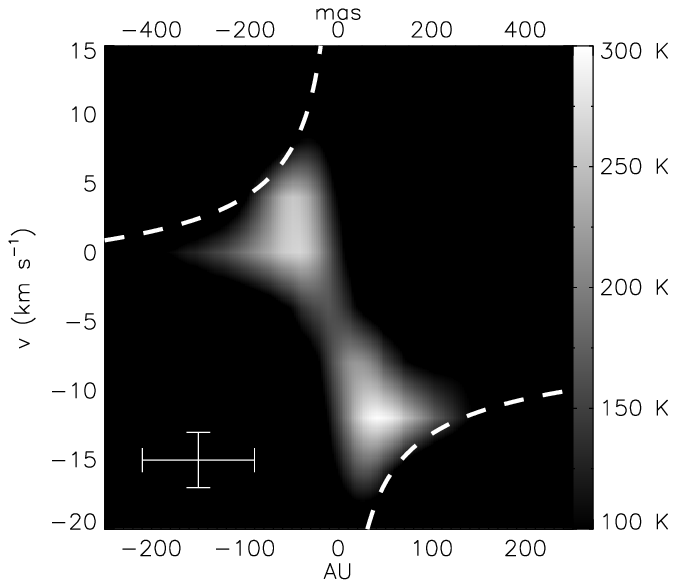}
\end{center}
\caption{
\label{diskobs}
The plots, adapted from \citet{krumholz07d}, show simulated observations of a massive protostellar disk in the CH$_3$CN $220.7472$ GHz line using ALMA (\textit{left}) and of the inner part of the same disk using the NH$_3$ $26.5190$ GHz line (the $(8,8)$ inversion transition) using the EVLA (\textit{right}). The left image shows velocity integrated brightness temperature as a function of position, while the right image shows brightness temperature as a function of position and velocity for a line along the disk plane. The beam sizes are indicated in both plots, and the color scales are chosen so that their bottom values correspond to the $3\sigma$ sensitivity of the assumed observation. The dashed line shows a Keplerian rotation curve for an $8.3$ $\msun$ star, the mass of the star at the center of the disk. Note the presence of very strong spiral structure in the position-position plot, and that the disk is offset from zero velocity in the position-velocity plot. Both of these are observational signposts of strong gravitational instability in the disk.
}
\end{figure}

Massive, gravitationally unstable disks are potentially observable, and their properties could be an important signpost of the massive star formation process. Simulated observations by \citet{krumholz07d} indicate that these disks should be visible out to distances of $\sim 0.5$ kpc using EVLA observations of molecular lines in the $20-30$ GHz range, and out to several kpc using ALMA observations of lines in the $200-300$ GHz range. Figure \ref{diskobs} illustrates some simulated observations of massive disks. Dust continuum observations, which provide morphological but not kinematic information, should be possible at considerably larger distances. These observations should reveal several indicators of gravitational instability, including strong spiral arms, hot spots produced by accreting protostars formed in the disk, rotation profiles that are super-Keplerian in their outer parts relative to their inner parts (due to the non-trivial disk to star mass ratio), and velocity offsets between the disk and star (due to motion of the star under the influence of the disk's gravity).

\section{The Competitive Accretion Model}
\label{compaccretion}

The primary difference between the core accretion model and competitive accretion is the
initial conditions and the physical process invoked to gather the mass of the most massive star. 
 Models of competitive accretion are derived from the requirement to form a full cluster of stars
in addition to forming the few massive stars in the system \citep{klessen98,bate03, bonnell03}.
As such, the initial conditions are dynamical unrelaxed allowing for large-scale
fragmentation into the many objects needsed to populate a stellar cluster. 
In this scenario, fragmentation produces lower mass stars while the few
higher-mass stars form subsequently due to  continued accretion  in stellar clusters
where the overall system potential funnels gas down to the centre of the potential, to be accreted by the proto-massive stars located there \citep[e.g.][]{bonnell01a, bonnell06c}. This model not only produces massive stars, but
a full IMF and mass segregated stellar clusters \citep{bonnell07a}.

Competitive accretion relies
primarily on the physics of gravity and that massive stars form in clustered environments.
That gravity is the dominant physical process in forming massive stars is not surprising as gravity need play a role
in both the larger scale formation of  stellar cluster as well as the smaller scale fragmentation
and collapse of individual objects.

\subsection{Initial fragmentation}

The initial fragmentation stage of a stellar cluster in a turbulent cloud involves the rapid generation
of structure due to the turbulence, followed by the collapse of individual fragments at Jeans length separations ($\approx 0.05$ pc) in the cloud \citep{larson84}. Such fragmentation is unaffected by any heating
from other newly formed stars due to this relatively large separation.  
Numerical simulations have repeatedly shown
that the gravitational fragmentation of a turbulent medium results in a range of stellar masses
based around the thermal Jeans mass of the cloud \citep{klessen98, klessen00a, klessen00b, bate03, bonnell04, bate05}. In the context of cloud core mass functions that are similar to the stellar IMF, this can easily be understood as being due to the subfragmentation of 
the higher-mass cores while the lower-mass ones are never gravitationally bound \citep{clark04, clark06}. This is just what has recently been observed in the Pipe nebula \citep{lada07}. 

Thus, turbulence drives structures into the molecular
cloud while the thermal physics determines the fragmentation scale and thus the characteristic mass for star formation \citep{larson05, jappsen05, bonnell06d}. Such fragmentation does not produce
stars with masses some tens larger than the Jeans mass implying that massive stars need
to form through an alternative mechanism. Turbulent fragmentation is sometimes invoked
to explain the formation of high-mass fragments \citep{padoan01}, but such fragments need to be very widely
separated and cannot account for the clustered environments of massive star formation.

Once individual fragments have formed, they fall together to form small-N systems which grow through
accretion of gas and other fragments.  These systems merge with other small-N clusters
generating a hierarchical merger scenario for stellar cluster formation much as is invoked
in cosmology. A hierarchcial fragmentation process is most straightforward in order to explain the
origin of stellar clusters. The infalling gas is crucial in providing the material to form the
more massive stars in the clusters. 

\subsection{Accretion in a cluster potential}
\begin{figure}
\centerline{\includegraphics[scale=0.35]{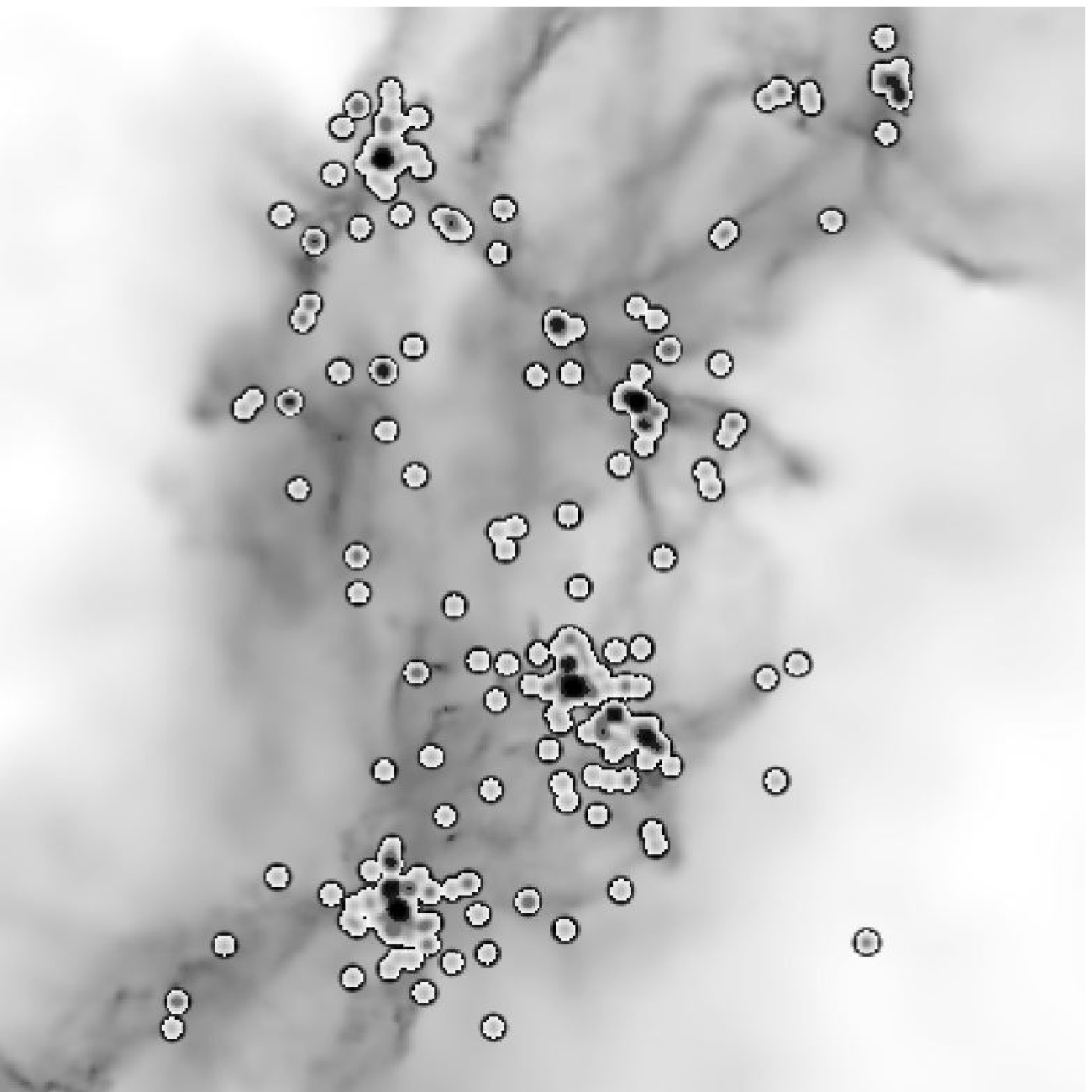}\includegraphics[scale=0.5]{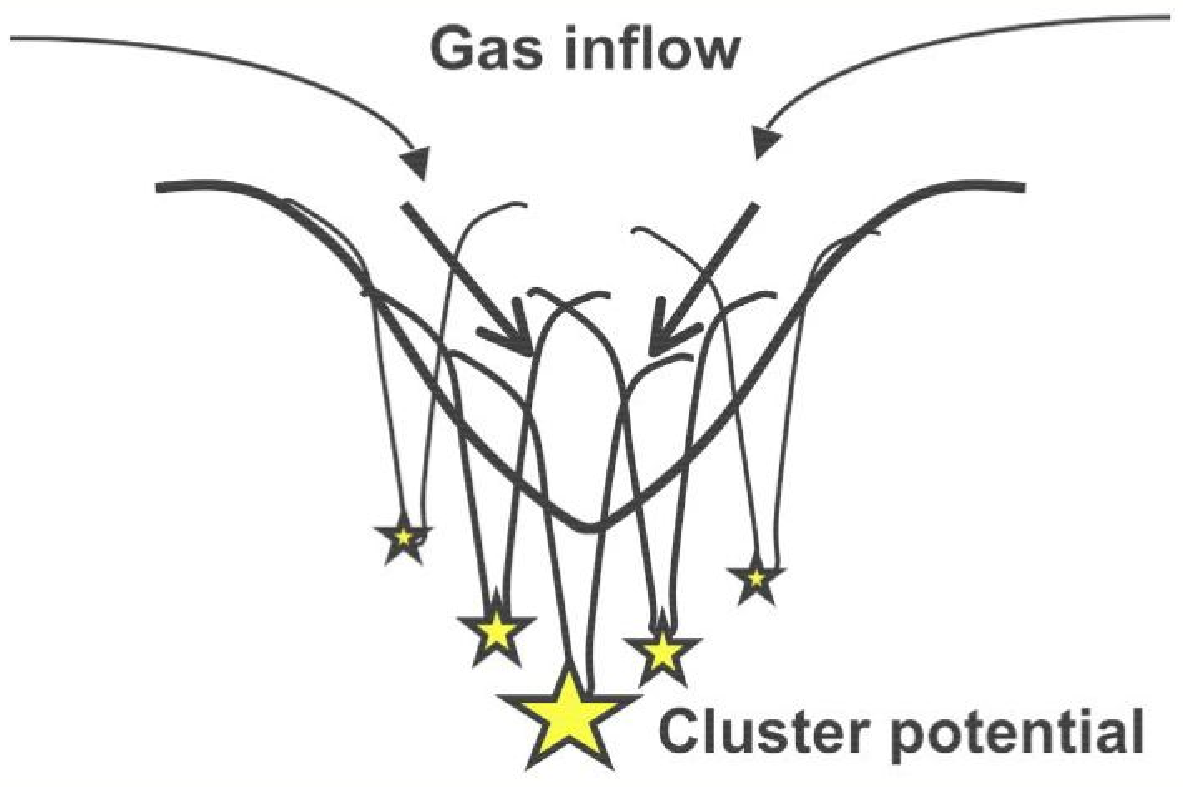}}
\caption{\label{compacc} The location of the most massive stars in a simulation of cluster formation (left)
shows their preferential location in the centres of clusters due to competitive accretion.
The right panel shows a schematic of the competitive accretion process whereby the 
that funnels gas down to the cluster core. The stars located there are therefore able
to accrete more gas and become higher-mass stars. The gas reservoir can be replenished by
infall into the large-scale cluster potential \citep[from][]{bonnell07a}.}
\end{figure}

Competitive accretion provides just such a mechanism as it relies on the continued accretion
onto a lower mass star due to the large-scale gravitational potential of a stellar cluster \citep{bonnell01a}. Such a potential naturally funnels gas down towards the centre of the system, there to be accreted by the soon to be massive star located there. 

Competitive accretion relies on  the inefficiency of fragmentation such that there is a large common reservoir
of gas from which the protostars can accrete \citep[e.g.][]{bonnell01a, bonnell04}. 
Observations of pre-stellar structures and of young stellar
clusters both support this view with the large majority of the total mass being in a distributed gaseous form \citep{motte98, johnstone00, johnstone04, lada03, bastian06}.
The second requirement is that the gas be free to move under the same gravitational acceleration as the stars. If the gas is fixed in place due to magnetic fields then accretion will be limited.
When these two requirements are filled, the dynamical timescale for accretion and evolution are similar such that a significant
amount of gas can be accreted. The necessary condition is that the system (cluster) is bound
such that the overall potential can have a significant effect. It is not necessary for the system
to be cold or collapsing as the potential will still funnel gas down to the centre even if the system is virialised.

\subsection{Dynamics of Accretion}

Competitive accretion can occur in two different regimes where the physics of the accretion depends
on the relative velocities between the stars and the surrounding gas.   The first occurs at the point
Firstly, at the point of fragmentation
the relative star-gas velocities are low (as are the star-star velocities). The accretion rates
are then determined by the tidal limits of each star's potential relative to the overall system potential.
This means that {\sl  the stars themselves do not need to move} relative to the surrounding gas. It is their
ability to gravitationally attract the gas, which is giverned by their tidal radius in the cluster potential, that determines the accretion rates and therefore the final mass distribution.

Secondly, once stars dominate the potential, they will virialise and have random velocities 
relative to the gas. Thus, in general, the relative velocities between gas and stars is large and the accretion is determined by a Bondi-Hoyle type accretion. This produces a steeper
mass spectrum due to the strong mass dependence of the accretion rates \citep{zinnecker84, bonnell01b}. Simulations and analytical arguments have shown that this type of process
results in a Salpeter-like IMF with mass spectra of $dN \propto m^{-\gamma} dm$ with $\gamma$ in the range of $2$ to $2.5$ \citep{bonnell01b, bonnell03}. It should be noted that even
in a virialised cluster their are exceptions to a Bondi-Hoyle accretion formalism. For example, the
most massive star sits in or near the centre of the potential. The relative velocity of the gas to this star
is then small  resulting in  near-spherical infall  limited more by the tidal radii of the cluster core.
Even stars that are further out and have higher velocities will periodically have velocities more aligned
to that of the infalling gas and thus a much lower relative velocity than would be naively predicted.

\subsection{Limitations of Competitive Accretion}
Recently, \citet{krumholz05e} have cast some doubt on this process by claiming that,
accretion in such environments cannot significantly increase a star's mass and therefore does not
play a role in establishing the stellar IMF.  In part this is correct as with competitive accretion,
most stars do not continue to accrete. It is the few that do that are important in terms of forming higher
mass stars and the IMF. What the Krumholz analysis overlooks is that Bondi-Hoyle accretion
\begin{equation}
\label{eqBH}
\macc \approx 4 \pi \rho \frac{\left( G \ms \right )^2}{v^3},
\end{equation}
is very dependent on the exact
values of the gas density$\rho$, stellar mass $\ms$, and velocity dispersion $v$, such that
with slight variations in these
properties, the accretion can vary from $10^{-9}$ \solmas\ yr$^{-1}$
to $10^{-4}$ \solmas\ yr$^{-1}$.  The higher values take more typical gas densities found
in the cores of stellar clusters and velocity dispersion of $0.5$ km s$^{-1}$,  in keeping with the
expectation of turbulence and the internal stellar velocity dispersion expected on these scales
\citep{bonnell06c}. 

The large range in accretion rates due to small differences in the gas density, the relative velocity and the initial stellar (fragment) mass highlight the strength of competitive accretion in explaining the 
full stellar IMF. It is only the rare stars that gain sufficient mass due to competitive accretion 
that attain high-mass status. As importantly, competitive accretion limits the growth of the bulk of the stars
keeping a low median stellar mass.

\subsection{Predictions}

\begin{figure}
\centerline{\includegraphics[scale=0.4]{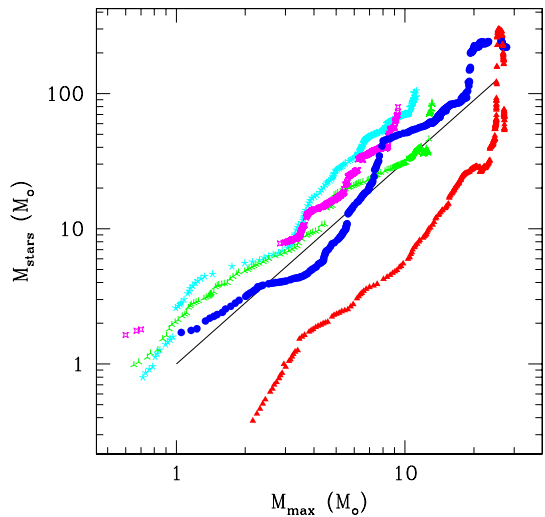}}
\caption{ The  total mass in stars is plotted against the mass of the most massive star in the system.
Competitive accretion predicts that as the total mass grows, the mass of the most massive star
also increases as $M_{max} \propto M_{stars}^{2/3}$  \citep{bonnell04}.\label{MmaxNcomp}}
\end{figure}

There are several observational predictions that can be gathered from the numerical simulations
of competitive accretion. The most striking of these is the relationship between the cluster
properties and the mass of the most massive star. Stellar clusters form when newly
formed stars fall together to produce small-N groups. These groups grow to form larger clusters
by accreting gas and stars from the surrounding environment. The gas accompanies the stars
into the cluster potential with a significant fraction being accreted by the more massive stars.
This naturally produces a correlation between the number of stars in the cluster (or the total mass in stars) and the mass of the most massive star therein \citep{bonnell04}. This prediction, where
\begin{equation}
M_{max} \propto M_{cluster}^{2/3},
\end{equation}
agrees remarkably well with the analysis of observed young stellar clusters by \citet{weidner06}.

Competitive accretion also provides an explanation for the frequency and properties
of binary systems among massive stars \citep{bonnell05}. Binary systems form readily through fragmentation
and  three-body capture in the small-N systems in the centre of the forming clusters. 
The continued accretion hardens  binary systems
while increasing the masses of the component. Thus, wide (100 AU) low-mass binary systems that
undergo continued accretion evolve into high-mass close binary systems. The frequency of such
high-mass binary systems is 100 per cent in these systems with the majority if these
being close with massive companions. They generally are found to have fairly high eccentricities which,
along with their small separations, implies that some of these should merge to form higher mass stars.
Stellar densities required to perturb such a binary
and force it to merge are estimated to be of order $10^6$ stars pc$^{-3}$. 

\subsection{Observational comparisons}

Competitive accretion requires a bound stellar system in order for the large-scale potential
to funnel the gas down to the forming massive stars. This does not mean that the system is cold
and globally collapsing, and in fact the system will continually appear to be nearly virialised
as any contraction will engender more transverse motions in the turbulent gas. 
On small scales,  the initial fragmentation  produces cores which are internally
subsonic as is expected in any turbulent cloud \citep{padoan01} as they form
due to the shock removal of the turbulent motions \citep{padoan01}. They also have a low velocity
relative to each other and to the nearby interfragment gas \citep{ayliffe07}. Once they collapse
into stars and move due to the large-scale potential, they gain significant velocity dispersions.

The fact that the accreting stellar clusters are always quasi-virial means that the star formation
process occurs over several free-fall times of the original system \citep{bonnell04}. This
is a lower limit as it neglects the formation time of the system which must be larger than
its localised free-fall time. A further complication to this arises from using final
dynamical times to compute the star formation rate. In \citet{bonnell03}, the final cluster
dynamical time is of order 3 times smaller than the initial crossing time of the cloud.
Using the final cluster timescale thus underestimates the star formation rate per free-fall
 time by at a factor of 3 or more such that the ONC estimates are very compatible
 to those from the simulation \citep{krumholz07e}. On large scales the SFR is lower still
 as much of the gas on such scales need not be bound \citep{clark05,clark07} .
 
\section{Challenges in Both Models: Feedback}
\label{challenges}

While the core accretion and competitive accretion models differ on how mass is gathered, they both face similar problems in explaining how that mass, once it reaches the vicinity of the star, gets to the stellar surface. The primary barrier is the radiation of the massive protostar, which will be extremely powerful because massive stars have very short Kelvin-Helmholtz times that allow them to reach the main sequence while still accreting. For a 50 $\msun$ ZAMS star, $t_{\rm KH}=GM^2/(RL)\approx 2 \times 10^4$ yr (using the ZAMS models of \citealt{tout96}), while the formation time such such a star is a $\mbox{few} \times 10^5$ yr \citep{mckee02}. Radiation can interfere with accretion both directly via the force it exerts and indirectly by ionizing the gas and raising its temperature above the escape speed from the star-forming region.

\subsection{Radiation Pressure}

The problem of radiation pressure in massive star formation can be viewed in terms of the Eddington limit. In spherical symmetry, a star of luminosity $L$ and mass $M$ shining on a gas of opacity $\kappa$ at a distance $r$ exerts a force per unit mass $\kappa L/(4\pi r^2 c)$, while the gravitational force per unit mass is $GM/r^2$. The gravitational force will be stronger only if $L/M < 4\pi G c/\kappa = 1.3\times 10^4 \kappa_1 (\lsun/\msun)$, where $\kappa_1$ is $\kappa$ measured in units of cm$^2$ g$^{-1}$. For ZAMS stars, the light to mass ratio reaches $10^4 \lsun/\msun$ by a mass of $35-40 \msun$, and the opacity of dusty gas at solar metallicity varies from hundreds at visible and UV wavelengths to a few in the far-infrared. This would seem to imply that accretion of dusty gas onto a protostar should be impossible due to radiation pressure for stars larger than a few tens of solar masses, a point made by a series of one-dimensional calculations \citep{larson71, kahn74, wolfire87}. 

Since more massive stars are observed to exist, something in this argument must fail. One possibility is that massive stars form not by accretion but by mergers of lower mass stars, which will not be affected by radiation pressure \citep{bonnell98, bonnell05}. However, while this process may take place, the high stellar densities required to produce an appreciable merger rate suggest that 
it is unlikely to be
the dominant formation mechanism under Galactic conditions.

A more likely route around the problem of radiation pressure is to recognize that, in an extremely optically thick protostellar envelope, the gas can reshape the radiation field by beaming it preferentially in certain directions. There is a certain subtlety to this point, since non-spherical \textit{accretion} is not by itself sufficient to overcome radiation pressure. If accretion were non-spherical but the radiation field remained spherical, then the net force on matter would still be away from the star at all points. However, if the opaque gas can force the radiation to flow around certain structures and travel away from the star non-spherically, then the radiation flux within those structures will be reduced and the net force can be inward, allowing accretion.

The most obvious candidate for a structure that can exclude the radiation field is a protostellar disk \citep{nakano89, nakano95, nakano00, jijina96}. In such a disk, direct stellar radiation will be absorbed in a thin layer within which the dust is near its sublimation temperature, roughly $1000-1500$ K depending on the exact composition of the grains \citep{pollack94, ossenkopf94}. Inside this layer, the opacity of the gas is small and radiation streams outward freely and spherically without exerting much force on the infalling matter. At the dust destruction front, radiation is rapidly down-converted to infrared wavelengths, and then diffuses outward through the dusty envelope. Since the disk is much more opaque than its surroundings, this diffusion is anisotropic, and much less flux passes through the disk than through the region around it. Matter that reaches the disk is shielded and is able to accrete. It will briefly feel a net outward acceleration as it passes through the dust destruction front and encounters the spherical radiation field within it, but if it has enough momentum inward its flow will not be reversed before the dust sublimes and it ceases feeling much radiative acceleration \citep{wolfire87}. Two-dimensional multi-group radiation-hydrodynamic simulations of this process starting from non-turbulent initial conditions demonstrate that it allows the formation of stars up to $43$ $\msun$ \citep{yorke02}. The upper mass limit is set in this simulation because eventually the radiation field is able to reverse the infall of matter which has not yet reached the disk, choking off further inflow to the disk.

Expanding to a three-dimensional view and including additional physical processes both open up additional possibilities for anisotropizing the radiation field. As happens in two dimensions, three-dimensional simulations show that radiation blows bubbles above and below the accretion disk, preventing gas from falling onto the star except through the disk. However, in three dimensions these bubbles prove to be subject to an instability analogous to Rayleigh-Taylor instability, which causes the bubbles to break up into low density chimneys filled with outward-moving radiation and high density walls that channel gas inward to the disk \citep{krumholz05d}. If there are substantial magnetic fields present in the infalling gas, this separation of the gas into dense inflowing channels and diffuse radiation chimneys can be significantly accelerated by photon bubble instability \citep{turner07}. Finally, both observations \citep{beuther02a, beuther03, beuther04, beuther05} and simulations \citep{banerjee07a} show that massive protostars produce hydromagnetic outflows just like their low mass counterparts. These outflows are launched near the star, where their dust is sublimed away, and the dust does not have time to re-form until the outflowing gas is well away from the star. As a result, the cavities that protostellar outflows punch into massive cores are optically thin, providing an escape channel for radiation that provides significant anisotropy and reduces the radiation pressure force near the equatorial plan by up to an order of magnitude \citep{krumholz05a}.

Together, these calculations suggest that the radiation pressure problem is considerably less daunting than was suggested by early spherically-symmetric estimates. However, this problem is not completely closed, and a number of outstanding issues remain. First, all of the multi-dimensional simulations and analytic calculations to date have assumed that dust and gas are perfectly well-coupled and that, except for subliming at high temperatures, grains do not change their properties as gas accretes. However, one-dimensional radiation-hydrodynamic simulations show that neither of these assumptions are strictly true \citep{suttner99}, and it is not clear how this will affect the radiation pressure problem in more dimensions.

Second, how magnetic fields will affect the radiation pressure problem is still largely an open question. Only one radiation-magnetohydrodynamic simulation of massive star formation has been reported in the literature \citep{turner07}, and this focused solely on magnetic effects in a small two-dimensional patch of a protostellar envelope. The role of magnetic fields on larger scales and in three dimensions remains unexplored.

Third, essentially all work on radiation pressure to date has taken place in the context of the core accretion rather than the competitive accretion model. The initial conditions generally start from a single gravitationally bound massive core. The sole exception to this is \citet{edgar04}, who simulate Bondi-Hoyle accretion onto a massive star and find that radiation pressure completely halts accretion onto stars larger than about $10$ $\msun$. Since the calculations begin with a smooth flow that is not undergoing gravitational collapse, it is unclear if the results will continue to hold in the context of  a more sophisticated competitive accretion model. However, this result does point out that whether radiation pressure can be overcome in the absence of an initial bound core remains an open question.

\subsection{Ionization}

The second form of feedback that might prevent accretion is ionizing radiation, which can heat gas to temperatures of $10^4$ K, with sound speeds of $10$ km s$^{-1}$. If the ionized region expands to include gas for which the escape speed from the massive star and its surrounding gas is smaller than $10$ km s$^{-1}$, the ionized gas will begin to escape, potentially choking off the supply of mass to the protostar. Whether or not this happens depends on the interaction between the ionizing radiation field and the gas inflow. \citet{walmsley95} first pointed out that in a spherically symmetric accretion flow, ionizing radiation can be trapped against the stellar surface simply by the flux of inflowing matter. The mass flux required to accomplish this is relatively modest, $\la 10^{-4}$ $\msun$ yr$^{-1}$ even for mid-O stars. The accretion rates for such massive objects expected from analytic models and simulations \citep[e.g.][]{mckee03, bonnell04, krumholz07a}, or simply inferred on dimensional grounds from the observed properties of massive star-forming regions, are of comparable size.

\citet{keto02, keto03, keto07} has generalized this model to cases where the accretion flow does not completely quench the HII region, and where the accretion flow is cylindrically-symmetric and has non-zero angular momentum. He finds that even after an HII region is able to lift off a stellar surface it may be confined for an extended period because the continual supply of new gas keeps the ionized region confined close enough to the star so that even gas with a sound speed of $10$ km s$^{-1}$ remains bound and accretes onto the star. In this case ionization continues despite the presence of an HII region. Depending on the relative sizes of the ionization radius and the gas circularization radius, some of the accretion disk may fall within the ionized region, leading to formation of an ionized, accreting disk as observed in G10.6-0.4 \citep{keto06b}. The general conclusion of this work is the ionizing radiation cannot significantly inhibit the formation of a massive star. As long as their is sufficient material left to accrete, the ionized region can be trapped so that gas can continue to accrete through it.

On the numerical side, \citet{dale05} has simulated the expansion of an HII region into a turbulent, non-magnetized parsec-sized cluster of gas and stars produced by a previous simulation of cluster formation via competitive accretion \citep{bonnell02}. In agreement with \citeauthor{keto02}, \citeauthor{dale05} find that ionizing radiation feedback cannot significantly impede the formation of massive stars. Although an ionizing source deposits in a protocluster gas cloud an amount of energy significantly larger than its binding energy, this energy does not succeed in unbinding the cluster or preventing the mass in dense structures from accreting. Instead, it simply blows off a relatively small fraction of the low density gas. Overall, the behavior of the gas with respect to the ionizing radiation has much in common with its behavior with respect to the non-ionizing radiation: an inhomogeneous gas flow shapes the radiation field, forcing radiation to flow out through low density channels while removing relatively little mass, and at the same time allowing dense, shielded gas to move inward under the influence of gravity.

Based on this combination of numerical and analytic work, it seems unlikely the ionizing radiation poses a serious challenge to either the core accretion or competitive accretion models of massive star formation.

\section{Future Directions}
\label{future}

\subsection{Observations}

There is considerable room for future observations to improve our models of how massive stars work. Such measurements are difficult to use directly in discriminating between the competitive and core accretion models, because the models differ primarily in how the mass is gathered, a process that occurs on time scales far too long to observe. The challenge, therefore, is to design observations that can address the secondary conditions implied or required by the models. Here we present some potentially powerful observational tests for this purpose. Most of these are submillimeter measurements that will require ALMA, although some may be possible earlier using the current generation of telescopes, such as PdBI, SMA, and CARMA. Others involve observations in the radio, infrared, and X-rays. These observations can be roughly divided into three scales: the ``micro" scale of individual star systems and their immediate gaseous progenitors, the ``meso" scale of individual star clusters, and the ``macro" scale of star formation on galactic scales.

\subsubsection{Micro Scales}

On the micro scale, one significant advance could come through observations of a large sample of disks around massive protostars. Both competitive accretion and core accretion models predict that massive stars should have disks $\sim 1000$ AU in size. However, competitive accretion models also predict that the typical massive star has a large number of close encounters with other stars which will destroy or greatly truncate the disk at regular intervals \citep{scally01, bonnell03, pfalzner06, moeckel07}. Continuing accretion causes the disk to grow back, but this model implies that in a sample of massive, embedded stars without significant HII regions, there should be some fraction at any given time that lack disks. In the core accretion model encounters are much rarer, and a disk should always be present until accretion stops and the remnant disk material is removed by ionizing radiation from the massive star. The fraction of diskless, still-embedded massive stars predicted by competitive accretion models is not yet certain, but an observation of whether such stars exist is a potentially powerful discriminator between models.

A second very useful micro scale observation would be a better determination of the properties of massive cores, the basic building blocks of the core accretion model. This model requires the similarity between the shapes of the core and initial stellar mass and spatial distributions observed for lower mass cores continue up to a least a few hundred $\msun$. Agreement between these properties has only been established in nearby low- and intermediate-mass regions, and evidence that this mapping does not hold for massive cores in more distant regions would rule out the simple core accretion model.

We would also like to know centrally concentrated massive cores are, and to what extent do they break up into collections of lower mass clumps at high resolution, since both of these will have an impact on how and whether they fragment. Observations today hint at a high degree of central concentration \citep{beuther07b}, but this is only for a single core, and even in this case the observations lack the resolution to address the degree of clumping. 

The question of clumping also leads to the problem of determining to what extent massive cores already contain low mass stars. \citet{motte07} find that even infrared-quiet massive cores harbor SiO$(2\rightarrow 1)$ emission indicative of outflows from low mass stars. They infer that any truly starless phase for massive cores must last $\la 1000$ yr. Given these cores' masses and mean densities, this is not surprising, since any reasonable density distribution would suggest that they contain at least $\sim \msun$ of gas at densities high enough to collapse in $1000$ yr. However, the degree of low mass stellar content is an important discriminator between theories. Core accretion models predict that massive cores fragment weakly, so one should essentially never find a massive core that has converted a significant fraction of its mass to stars but has formed only low mass stars. Any low mass stars that exist in such a core should grow to become massive stars as more gas falls onto them. Competitive accretion models, on the other hand, predict that an outcome in which massive cores fragment and produce only low mass stars should not be uncommon. For this reason, determining the low-mass stellar content of IR-quiet massive cores would be extremely useful. This might be possible using X-ray observations, which are very sensitive to the presence of low mass stars \citep[e.g.][]{feigelson05}.

A final micro scale observational question concerns the apparently isolated O stars identified by \citet{dewit04, dewit05}. The existence of these stars already seems to suggest that the presence of a cluster is not absolutely required to form massive stars. It would be very helpful to know if these stars are unusual in any way other than being isolated. For example, are their rotation rates of magnetic properties different than those of O stars found in clustered environments? This would help address the question of whether these stars formed by a different process than typical O stars.

\subsubsection{Meso Scales}

On the meso scale, the most useful observations would be better constraints on the dynamical state of cluster-forming gas clumps, and on the formation history of the star clusters that result. As discussed earlier, the question of how massive stars form is intimately tied to the question whether star clusters form in a process of global collapse in which the parent cloud never reaches any sort of equilibrium configuration, or whether cluster-forming clouds are quasi-equilibrium objects within which collapse occurs only in localized regions. The data that exist now can be interpreted in conflicting ways \citep[e.g.][]{elmegreen00, tan06a, krumholz07e, elmegreen07, huff07}, and cannot definitively settle which model more closely approximates reality.

For gaseous star-forming clumps, one can approach this problem by searching for signatures of supersonic collapse via line profiles. There is some evidence for such collapse in certain regions of star cluster formation \citep[e.g.][]{peretto06}, but in others infall is either completely absent or is subsonic \citep[e.g.][]{garay05}. The question of whether the typical region of massive star formation is in a state of global collapse, however, cannot be answered without more systematic searches covering large number of star-forming clouds.

For revealed star clusters, observations can help by pinning down the star formation history in more detail, hopefully settling the question of the true age spreads in clusters.  This project will require both improvements in observational data and in the pre-main sequence models used to convert observed stellar luminosities and temperatures into ages. Observationally, the current surveys suffer from potential problems with incompleteness, interlopers, and dust obscuration. Incompleteness preferentially removes the oldest and lowest mass members from a sample, making clusters appear younger than their true age, while interlopers tend to be older stars whose inclusion in the sample can artificially age clusters. Obscuration by dust in the vicinity of the cluster can push the age in either direction, since it worsens incompleteness but, by dimming stars, it also makes them appear older than their true age. Future surveys should be able to do considerably better than the current generation in minimizing these effects.

For the pre-main sequence tracks, improvements are likely to come from empirical calibrations of the models against binary systems, where masses can be obtained dynamically and the stars are likely to be coeval \citep[e.g.][]{palla01,stassun04,boden05}. As more pre-main sequence binaries are discovered and their masses and radii determined, it should be possible to progressively improve the models and reduce the uncertainties in the ages they produce.

A final point to remember is that distance uncertainty is a constant concern, since, as the case of the ONC shows, even a $\la 20\%$ change in the estimate distance can significantly skew estimates of the star formation history. Radio parallax measurements of the distances to many more star forming regions should be forthcoming in the next few years, and may force significant revisions in our models of star formation histories.

\subsubsection{Macro Scales}

Galactic and larger scale observations are powerful tools for massive star formation studies because the offer the opportunity to obtain statistical constraints, something generally impossible on smaller scales due to the rarity of massive stars. Such observations already indicate the existence of an upper mass cutoff of $\sim 150$ $\msun$, beyond which stars appear to be very rare or non-existent \citep{elmegreen00b, weidner04, figer05}. It is possible that this limit is a result of instabilities in stars after they form rather than something set at formation, but if it is set at formation then it is a result that massive star formation models should be able to explain. An observational determination of there is any variation of this maximum mass from galaxy to galaxy, or within our galaxy, might provide a valuable clue to the origin of the cutoff and thus to the correct picture of how massive stars from. More generally, any convincing evidence for variation in the IMF from region to region, beyond that expected simply from statistical fluctuations, would be helpful.

Galactic-scale observations can also be useful in constraining the star formation rate as a function of density. \citet{krumholz05c} define the star formation rate per free-fall time, $\sfrff$, as the fraction of a gaseous object's mass that is incorporated into stars per free-fall time of that object. \citet{krumholz07e} show that, when averaged over galactic scales, $\sfrff$ is typically a few percent for objects ranging in density from entire giant molecular clouds, $\sim 10^2$ cm$^{-3}$ in density, to sub-pc size clumps traced by HCN$(1\rightarrow 0)$ emission, which is excited only at densities $\sim 10^5$ cm$^{-3}$ or higher. They argue that this implies that cluster-forming gas clumps cannot be rapidly collapsing, a result required by the core accretion model and incompatible with competitive accretion. However, systematic uncertainties limit the determination of $\sfrff$ to factors of $\sim 3-4$, meaning that a rise of $\sfrff$ with density cannot be definitively ruled out. The largest such uncertainty is in the conversion of observed molecular line luminosities to masses, and this can be significantly reduced by better observational calibrations of the ``X" factor for the various molecules used in surveys.

Observations to date also cannot rule out the possibility that the galactic-averaged value of $\sfrff$ is low because it averages over a significant amount of the emitting gas is not associated with star-forming regions. In this picture star-forming clumps would collapse rapidly, as required by competitive accretion, but their high values of $\sfrff$ would be diluted by the inclusion in galactic averages of a significant amount of gas that emits in a given line but that is not bound and is not forming stars at all. Resolved observations of a large sample of line-emitting regions in nearby galaxies, combined with resolved infrared observations to estimate star formation rates in those regions, could determine whether such sterile clouds exist. A preliminary survey of the Milky Way finds that large HCN-emitting gas clouds generally have the same values of $\sfrff$ inferred from galactic averages \citep{wu05}, but there is considerable scatter in the result, and since the sample was selected based on indicators of massive star formation, the survey does not address the possible existence of sterile HCN-emitting clouds.

\subsection{Simulations}

There is also considerable room for improvement in simulations of massive star formation. The simulations that have been performed to date can be roughly placed into a continuum running from those with large dynamic ranges in three dimensions but rather limited inputs physics to those that include extremely detailed physics, but with lower dimensionality, lower dynamic range, or both. Including all or even most of the physical processes relevant to massive star formation in a simulation that has the same dynamic range as that achieved by calculations including only hydrodynamics plus gravity is not feasible on current computers. However, it is possible to push from both ends of the spectrum toward the middle. We end this review by highlighting what we believe are two of the most promising directions for progress of this sort.

\subsubsection{Cluster Formation with Feedback}

Much of the debate between competing models concerns whether the collapse process that forms a massive star is localized to a single massive core, or occurs globally within a much larger cloud that forms an entire cluster. However, the dynamics of clusters are certainly modified by feedback from embedded protostars, and the role of feedback has been subjected only to a very limited exploration in simulations.

One obvious place for improvement would be to extend the work of \citet{krumholz07a} on radiative suppression of fragmentation to larger scales. The \citeauthor{krumholz07a} simulations follow the collapse of $\sim 0.1$ pc-size cores down to $\sim 10$ AU scales; the cores are turbulent initially, but the turbulence is allowed to decay freely. It should be possible to repeat a simulation of this sort on large scales, following an entire cluster-forming gas clump $\sim 0.5$ pc in size and $\sim 5000 \msun$ in mass. Such a simulation would address the question of whether the dynamics of competitive accretion, which inevitably happen in a simulation containing many thermal Jeans masses where the turbulence is not driven, are significantly modified by radiative feedback. The challenge here is whether the calculations could be done at high enough resolution to allow meaningful comparison to earlier work and still run in a reasonable amount of time. Simulations involving radiative transfer are generally $\sim 5$ times as expensive as ones involving only hydrodynamics and gravity, and their resolution is correspondingly worse. New radiation-hydrodynamic algorithms that are potentially much faster than those used in the \citeauthor{krumholz07a} simulations may help with this problem \citep[e.g.][]{whitehouse04, whitehouse05, shestakov07}.

Another important priority is to improve on the pioneering work of \citet{li06b} and \citet{nakamura07} on the effect of protostellar outflows on cluster dynamics. They simulate magnetized clouds in which forming protostars generate outflows, and find that these outflows drive turbulence strongly enough to keep the star formation rate to $\sfrff \sim 5\%$ and to prevent the cloud from going into overall collapse. If these results are valid, then outflows qualitatively change the dynamics of star cluster formation in a way that appears on its face inconsistent with the competitive accretion model in which massive stars form by accreting gas over large distances from a region in global rather than local collapse. However, there are significant caveats. The simulations use a fairly low resolution fixed grid with no adaptivity. Low resolution could artificially isotropize the energy from outflows, making them more effective at driving turbulence than more collimated outflows would be. The simulations also use a very simple implementation of sink particles to represent forming protostars; these sink particles all have the same mass, and they instantaneously appear on the computational grid rather than accreting gas over time. As a result, they cannot directly study the impact of outflows on the dynamics of accretion. Clearly it is important to repeat these simulations using higher resolution adaptive mesh or Lagrangean simulations, and using a more sophisticated implementation of sink particles in which it is possible to follow accretion over time \citep{bate95, krumholz04}.

\subsubsection{Fragmentation and Feedback with Magnetic Fields}

Another almost completely unexplored area is the effect of magnetic fields on fragmentation, accretion, and feedback in regions of massive star formation. The first MHD simulations of low mass star formation with enough dynamic range to study fragmentation, using three very different numerical techniques, all find that magnetic fields significantly reduce fragmentation compared to the purely hydrodynamical case \citep{hosking04, price07, hennebelle07b, machida07}. To date no simulations of magnetic fields in the context have massive star or star cluster formation have been performed. It would be very valuable to repeat earlier hydrodynamic or radiation-hydrodynamic simulations (e.g.\ \citealt{bonnell04} or \citealt{krumholz07a}) with MHD to see how magnetic fields change fragmentation and initial accretion.

Magnetic fields may also also change how feedback from embedded protostars interact with the environment, an effect already hinted at by a number of simulations. \citet{turner07} show that magnetic fields make it easier to overcome the radiation pressure barrier in massive star formation because they allow photon bubble instabilities to develop, while \citet{nakamura07} show that magnetic fields, by providing a means to redistributed energy from protostellar outflows throughout a cluster, enhance outflows' ability to drive turbulent motions and affect cloud dynamics. \citet{krumholz07f} find that magnetic fields can help confine HII regions and also transmit their effects via fast magnetosonic and Alfv\'{e}n waves to distant parts of a flow. This could potentially alter the findings of \citet{dale05} that because HII regions expand asymmetrically into low density parts of a flow they cannot unbind clusters or greatly reduce the star formation rate within them. In summary, all the simulations of massive star formation that have been performed to date could be very significantly altered by the inclusion of magnetic fields. With the advent of AMR and SPH MHD codes \citep{fromang06, price07} capable of achieving significant dynamic ranges in MHD problems, it is important to repeat these calculations with MHD.

\acknowledgements We thank S.~S.~R. Offner for helpful discussions. This work was supported by: NASA through Hubble Fellowship grant HSF-HF-01186 awarded by the Space Telescope Science Institute, which is operated by the Association of Universities for Research in Astronomy, Inc., for NASA, under contract NAS 5-26555; the National Science Foundation under Grant No. PHY05-51164; the Arctic Region Supercomputing Center; the National Energy Research Scientific Computing Center, which is supported by the Office of Science of the US Department of Energy under contract DE-AC03-76SF00098, through ERCAP grant 80325; the NSF San Diego Supercomputer Center through NPACI program grant UCB267; and the US Department of Energy at the Lawrence Livermore National Laboratory under contract W-7405-Eng-48.


\end{document}